\documentclass[a4paper,oneside,english]{amsart}
\usepackage{graphicx}
\usepackage{amssymb}

\makeatletter
\theoremstyle{plain}
\theoremstyle{plain}
\newtheorem{thm}{Theorem}
\theoremstyle{remark}
\newtheorem*{rem*}{Remark}

\newtheorem*{thm*}{Theorem}
\newtheorem{theorem}{Theorem}

\newtheorem{lemma}[theorem]{Lemma}

\newcommand{\be}{\begin{equation}}
\newcommand{\nd}{\noindent}
\newcommand{\ee}{\end{equation}}
\newcommand{\ben}{\begin{eqnarray}}
\newcommand{\een}{\end{eqnarray}}

\usepackage{babel}
\makeatother

\begin{document}
\title{Superstatistics based on  the microcanonical ensemble}
\author{{\it C. Vignat$\,^1$, A. Plastino$\,^2$,}  and {\it A. R. Plastino$\,^{2,3}$} \\\\ $^1$ L.I.S. Grenoble, France and E.E.C.S., University of Michigan, USA\\ e-mail: vignat@univ-mlv.fr  \\\\ $^2$ La Plata National University-CONICET  \\ C. C. 727 - 1900 La Plata - Argentina; e-mail: plastino@fisica.unlp.edu.ar \\\\ $^3$  Department of Physics, University of Pretoria, Pretoria 0002, South Africa; e-mail: arplastino@maple.up.ac.za}

\begin{abstract}
Superstatistics is a ``statistics" of ``canonical-ensemble statistics". In analogy, we consider here a similar theoretical construct, but
based upon the microcanonical ensemble. The mixing parameter is not the temperature but the index $q$ associated with
the non-extensive, power law entropy $S_q$. \vskip 2 mm
 \nd {\sf Keywords:} Superstatistics,  Tsallis entropies \vskip 2mm \nd
{\sf PACS:} 05.40.-a, 05.20.Gg
\end{abstract}

\maketitle

\section{Introduction}
\nd The  most notorious and renowned probability distribution (PD) used in the field of statistical mechanics is undoubtedly that deduced by Gibbs for the canonical ensemble \cite{reif, pathria}, usually referred to as the Boltzmann-Gibbs (BG) equilibrium distribution
\be \label{gibbs}  p_G(i) =\frac{\exp{(-\beta E_i)}}{Z_{BG}}, \ee with $E_i$ the  energy of the microstate labeled by $i$, $\beta=1/k_B T$ the inverse temperature ($T$), $k_B$ Boltzmann's constant, and $Z_{BG}$ the partition function. The exponential term
$F_{BG}=\exp{(-\beta E)}$ is, of course, called the BG factor. Recently Beck and Cohen
\cite{BC03,B04a} have advanced a generalization
(called ``superstatistics")  of this BG factor, assuming that
the inverse temperature $\beta$ is a
stochastic variable. They effect a multiplicative convolution
to obtain a generalized statistical factor
$F_{GS}$ in the fashion \cite{B04a}
\be \label{bc} F_{GS}=\int_0^\infty\, d\beta\, \beta f(\beta)\, \exp{(-\beta E)}\equiv f \circ F_{BG},\ee
where $f(\beta)$ satisfies
\be \label{normita} \int_{-\infty}^{\infty}\, d\beta \, f(\beta) =1.\ee
Note that the $\circ-$sign is used to denote the multiplicative convolution between two PDs: multiplicative convolution of two PDs $f_{X}$ and $f_{Y}$ is the PD of the product of the two corresponding random variables $X$ and $Y$.
Superstatistics would arise as a ``statistics of statistics" that
takes into account fluctuations temperature
\cite{BC03,B04a,B01,BC04,B04b,B04c,TB05}.
Fixed temperature values, as those
emblematic of the canonical ensemble of Gibbs',
require that the system to be described be in thermal contact,
and in equilibrium,
with an infinite heat reservoir characterized precisely by that temperature $T$.

\vskip 3mm \nd Beck and Cohen also show that, if $f(\beta)$ is a
$\chi^2$ distribution, nonextensive thermostatistics (NEXT) is
obtained, which is of interest because NEXT, also known as
Tsallis' thermostatistics, is today a very active field, perhaps
a new paradigm for statistical mechanics, with applications to
several scientific disciplines \cite{gellmann,lissia,fromgibbs}.
In working in a NEXT framework  one has to deal with power-law
distributions, which are certainly ubiquitous in physics (critical
phenomena are just  a conspicuous example \cite{goldenfeld}).
Indeed, it is well known that  power-law distributions arise quite
naturally in maximizing Tsallis'  information measure   ($q$ is a
real parameter called the ``nonextensivity index") \be H_{q}\left(
f\right)  =\frac{1}{1-q}\left(  1-\int_{-\infty}^{+\infty}f(x)^{q}
dx\right), \label{dino} \ee subject to appropriate constraints. In
the case of the canonical distribution there is only one
constraint, the energy $E$, i.e.,  $\langle E\rangle = K\,\,( {\rm
K\,\,a\,\,\, constant})$ and  the equilibrium canonical
distribution writes  $f(x) =(1/Z_q)[1-(1-q)\beta_q
E]^{\frac{1}{q-1}}$, with $\beta_q$ and  $Z_q$ standing for the
NEXT counterparts of $\beta$ and $Z_{BG}$ above. It is a classical
result that as $q\rightarrow1,$   Tsallis entropy reduces to
Shannon entropy

\be \label{shannon}
H_{1}\left(  f\right)  =-\int_{-\infty}^{+\infty}f(x)\log f(x).
\ee

\nd Tsallis' PDs are encountered in analyzing a rather large variety of physical systems \cite{b1,b2,b3,b4,b5}, which encourages people to continue investigating the nonextensive formalism along multiple viewpoints  and a multitude of paths.

\vskip 4mm
\nd In such a spirit, and further pursuing along the road first travelled by
Beck and Cohen \cite{BC03}, we show here that a new type of superstatistics can be constructed for which
the fluctuating physical quantity is the nonextensivity index $q$.

\section{Mathematical framework}
\subsection{Notations}
\nd Using notation $x_{+}=\max(x,0)$, let us denote \be \label{1}
S_{q,\sigma^{2}}\left(x\right)=A_{q}\left(1-\frac{x^{2}}{\sigma^{2}}\right)_{+}^{\frac{1}{q-1}},\ee
the Tsallis distribution with variance $\sigma^{2}\frac{q-1}{3q-1},$ nonextensivity index  $q$,
and normalization constant \be \label{2}
A_{q}=
\frac{\Gamma\left(\frac{1}{q-1}+3/2\right)}
{\Gamma\left(\frac{1}{q-1}+1\right)\sigma\sqrt{\pi}}.\ee
A word of caution is necessary here. There are several versions of Tsallis'
thermostatistics. In some of them Tsallis' PDs carry the power
\ben \label{qestre}
  (i) & \,\, \frac{1}{1-q}\,
  \,{\rm while\,\,for\,\,others\,\, it\,\, is}
  \cr  (ii)   & \,\,   \frac{1}{q-1}\,
  \,{\rm  (our\,\, choice\,\, above)}. \een
Of course, translation between the two is straightforward.
If we denote (a) by $q^*$ the nonextensivity index of the first choice above, and
(b) by $q$ that of the second one,  then

\be \label{qestre1} q^*=2-q. \ee 
\subsection{Main result}
Before stating our main result under the form of Theorem \ref{thm1}, we need the following lemma.
\begin{lemma}
\label{lemma1}
\nd $\forall\sigma,$
$\forall a>0,$ there exists a discrete distribution $p_{k}\left(a\right)$
such that
\begin{equation}
\sum_{k=0}^{+\infty}p_{k}\left(a\right)S_{1+\frac{1}{k},2a\sigma^2}\left(x\right) = \begin{cases} \frac{1}{\text{erf}(\sqrt{a})\sigma\sqrt{2\pi}}\exp\left(-x^{2}/2\sigma^{2}\right) & \forall x\in\left[-\sigma\sqrt{2a},+\sigma\sqrt{2a}\right]  \\
0 \hspace{4cm} & {\rm else}
\end{cases}
\label{eq:expansion}
\end{equation}
\end{lemma}

\begin{proof}
\nd Assuming $x\in\left[-\sigma\sqrt{2a},+\sigma\sqrt{2a}\right]$, let us compute the left hand side of (\ref{eq:expansion}) as
\begin{eqnarray*}
\sum_{k=0}^{+\infty}p_{k}\left(a\right)\frac{1}{\sigma\sqrt{2a}}S_{1+\frac{1}{k},1}\left(\frac{x}{\sigma\sqrt{2a}}\right) 
& = & \frac{\exp(-a)}{\text{erf}(\sqrt{a})} 
\sum_{k=0}^{+\infty} \frac{a^{k+1/2}}{\Gamma(k+3/2)}\frac{\Gamma(k+3/2)}{\sigma \sqrt{2a} \Gamma(k+1) \sqrt{\pi}} \left(1-\frac{x^{2}}{2a\sigma^{2}}\right)_{+}^{k}\\
& = & \frac{\exp(-a)}{\text{erf}(\sqrt{a})} \frac{1}{\sigma \sqrt{2\pi}} \exp\left(a-\frac{x^{2}}{2\sigma^{2}}\right) \\
& = & \frac{1}{\text{erf}(\sqrt{a}) \sqrt{2\pi}}\exp\left(-x^{2}/2\sigma^{2}\right).
\end{eqnarray*}

\nd If $x\notin\left[-\sigma\sqrt{2a},+\sigma\sqrt{2a}\right]$,
 then $S_{1+\frac{1}{k},1}\left(\frac{x}{\sigma\sqrt{2a}}\right)=0 \  \forall k$, and,
 consequently,  the sum obviously vanishes.
 The fact that $p_{k}(a)$ is a distribution can be checked by integrating over $x\in\mathbb{R}$ both sides of relation (\ref{eq:expansion}): we obtain
\begin{equation}
\nonumber \sum_{k=0}^{+\infty}p_{k}\left(a\right)  =  \int_{-\sigma\sqrt{2a}}^{+\sigma\sqrt{2a}}
\frac{1}{\text{erf}\left(\sqrt{a}\right)\sigma\sqrt{2\pi}}\exp\left(-x^{2}/2\sigma^{2}\right) 
 = 1.
\end{equation}

\end{proof}

\nd Consider now a discrete
random variable $Q\left(\omega,a\right)$ such that \be \label{3}
\Pr\left(Q\left(\omega,a\right)=1+1/k\right)= p_{k}\left(a\right) =
\frac{\exp\left(-a\right)a^{k+\frac{1}{2}}}{                                                                                                                                                \Gamma\left(k+\frac{3}{2}\right) \text{erf}(\sqrt{a})},k\in\mathbb{N}.\ee

\begin{figure}[htbp]
\begin{center}
\centering
\includegraphics[width= 2.9776in, height= 2.6097in]{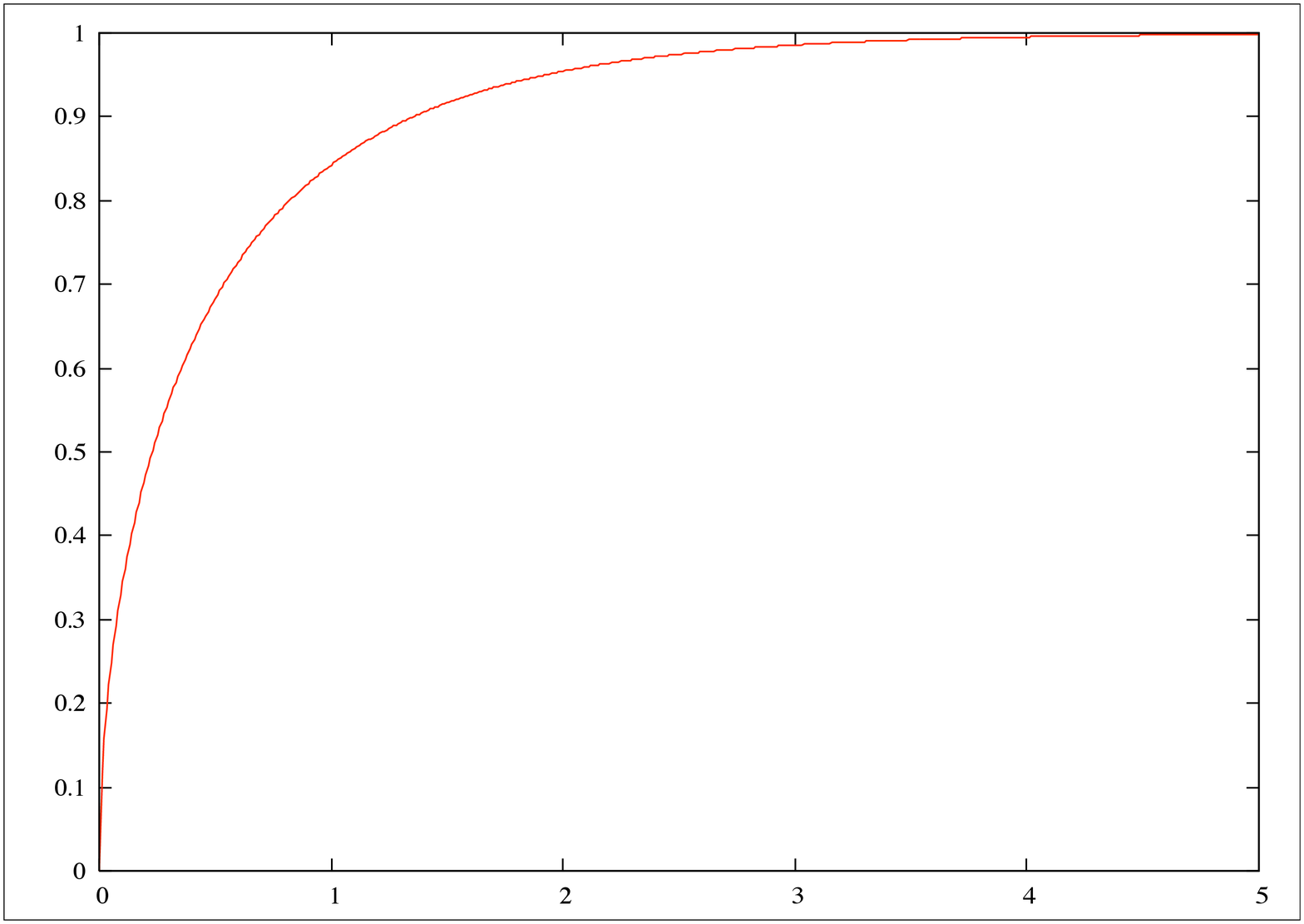}
\caption{function $f\left(a\right)=\text{erf}\left(\sqrt{a}\right)$}
\end{center}
\end{figure}

As shown on Fig. 1, function $f\left(a\right)=\text{erf}\left(\sqrt{a}\right)$ gets
rapidly close to 1 as $a$ increases - typically $f(5)=0.9984$ - so that 
\be
p_{k}(a)=\frac{\exp\left(-a\right)a^{k+1/2}}
{\Gamma\left(k+3/2\right)} 
\label{shiftedpoisson}
\ee 
can be considered as a very good approximation to the
distribution of $Q(\omega,a)$ for large values of $a$. Thus a stochastic expression of equality (\ref{eq:expansion}) for large values of $a$ writes as follows: 
\begin{thm} 
\label{thm1}
A Gaussian distribution with variance $\sigma^{2}$ can be accurately approximated - in the sense of lemma \ref{lemma1} - as a $q$-mixture of Tsallis maximizing distributions as defined by (\ref{1}) where random non-extensivity parameter $Q(\omega,a)$ distributed according to (\ref{shiftedpoisson}).
\end{thm}
Figure 2 shows the partial sums $\sum_{k=0}^{n}p_{k}(a)S_{1+\frac{1}{k},2a\sigma^2}\left(x\right)$ involved in formula (\ref{eq:expansion}) for $n=0$, $5$, $10$ and $\infty$ (Gaussian limit), in the case where $a=5$ and $\sigma=1$. We remark that the $k=0$ term in the sum corresponds to a uniform distribution on interval $\left[-\sigma\sqrt{2a},+\sigma\sqrt{2a}\right]$

\begin{figure}[htbp]
\begin{center}
\centering
\includegraphics[width= 2.9776in, height= 2.6097in]{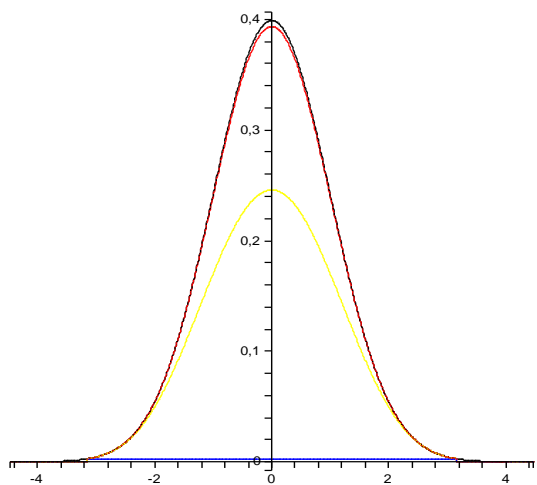}
\caption{values of $\sum_{k=0}^{n}p_{k}(a)S_{1+\frac{1}{k},2a\sigma^2}\left(x\right)$ for $a=5$, $\sigma=1$ and (from bottom to top) $n=0, 5, 10 \,\, \text{and} +\infty$}
\end{center}
\end{figure}

\subsection{The distribution of the non-extensivity parameter}
\nd We first remark that a classical Poisson distribution with parameter $\lambda$ writes
\be
\label{poisson}
q_{k}(\lambda)=\exp\left(-\lambda\right)\frac{\lambda^{k}}{k!},\ k\in\mathbb{N}.
\ee
The distribution $p_{k}\left(a\right)$, as written in (\ref{shiftedpoisson}),
can thus be considered as a ``shifted'' Poisson distribution.

\begin{thm}
\nd The expectation and variance of a random variable $X$ such that
\be
\nonumber
\Pr\{X=k\}=p_{k}(a)
\ee
can be approximated, for  large enough $a$, by
\[
m  =  a-\frac{1}{2},\,\,  
\sigma^{2}  = a
\]
\end{thm}

\begin{proof}
\nd The expectation writes
\begin{equation*}
m  =  \sum_{k=0}^{+\infty}kp_{k}\left(a\right)
 =  \frac{\exp\left(-a\right)}{\text{erf}(\sqrt{a})}\sqrt{\frac{a}{\pi}}+\left(a-\frac{1}{2}\right)
\end{equation*}
Thus for $a$ large, $m\simeq a-\frac{1}{2}.$ The variance is
\begin{eqnarray*}
\sigma^{2} & = & \sum_{k=0}^{+\infty}k^{2}p_{k}\left(a\right)-m^{2}\\
 & = & \frac{\exp\left(-a\right)}{\text{erf}\sqrt{a}}\sqrt{\frac{a}{\pi}}\left(a-\frac{1}{2}\right)+\left(a^{2}+\frac{1}{4}\right)-m^{2}\\
 & \simeq & a^{2}+\frac{1}{4}-\left(a-\frac{1}{2}\right)^{2}
 \end{eqnarray*}
so that, for large $a$, $\sigma^{2}\simeq a.$ 
\end{proof}

We note the resemblance of this result with the case of Poisson distribution (\ref{poisson}) for which
\[
m=\lambda,\,\,
\sigma^{2}=\lambda.\]

\subsection{Extension to multivariate and power exponential distributions}
\nd The preceding results can be generalized from univariate Gaussians to multivariate power exponential distributions as follows. First we recall the following result from
\cite{alf2}.
\begin{thm}
The maximizer of Tsallis entropy with $q>1$ under the constraints  \begin{equation}
\langle \vert X_{i} \vert ^{\gamma_{i}} \rangle \equiv \nonumber
E\vert X_{i} \vert ^{\gamma_{i}}=K_{i}\label{eq:constraints}\end{equation}
is
\begin{equation}
f_{X,q}\left(X\right)=\begin{cases}
A_{q}\left(1-\sum_{k=1}^{n}\lambda_{k} \vert x_{k} \vert ^{\gamma_{k}}\right)^{\frac{1}{q-1}} & x\in E_{1}\\
0 & else\end{cases}\label{eq:tsallis}
\end{equation}
where the hyperellipsoid $E_{1}$ is defined as \[
E_{1}=\left\{ X\in\mathbb{R}^{n}\vert\sum_{k=1}^{n}\lambda_{k}\vert x_{k} \vert ^{\gamma_{k}}\le1\right\} \]
and the Lagrange multipliers $\lambda_{k}$ are such that the constraints
(\ref{eq:constraints}) are verified: if $q>1,$ all
Lagrange multipliers $\lambda_{k}$ are positive. Moreover, constant
$A_{q}$ writes as follows:\[
A_{q}=\frac{\Gamma\left(\frac{1}{q-1}+\frac{1}{\gamma}+1\right)}{\Gamma\left(\frac{1}{q-1}+1\right)}\prod_{k=1}^{n}\frac{\gamma_{k}\lambda_{k}^{1/\gamma_{k}}}{2\Gamma\left(1/\gamma_{k}\right)}.\]
with $\frac{1}{\gamma}=\sum_{k=1}^{n}\frac{1}{\gamma_{k}}.$
\end{thm}

Now we are in position to state the following theorem.
\begin{thm}
\nd A generalized exponential distribution\[
g\left(x\right)=\left(\prod_{k=1}^{n}\frac{\gamma_{k}\lambda_{k}^{1/\gamma_{k}}}{2\Gamma\left(1/\gamma_{k}\right)}\right)\exp\left(-\sum_{k=1}^{n}\lambda_{k} \vert x_{k}\vert^{\gamma_{k}}\right)\]
$\left(\lambda_{k}>0\,\,\forall k\right)$ can be approximated as a $q-$mixture
of Tsallis maximizing distributions $f_{x,q}\left(x\right)$ as defined
by (\ref{eq:tsallis}), in the following sense: $\forall a>0,$
\[
f_{Ax,Q(\omega,a)}\left(x\right)=
\begin{cases}
B_{a,\gamma}g\left(x\right) & \forall x\in E_{a}\\
0 & else
\end{cases}
\]
where \[
A=diag\left\{ a_{k}^{1/\gamma_{k}}\right\} ,E_{a}=\left\{ y\in\mathbb{R}^{n}\vert\sum_{k=1}^{n}\lambda_{k} \vert y_{k} \vert ^{\gamma_{k}}\le a\right\} \]
and $Q(\omega,a)$ is a random nonextensivity parameter such that
\[
\Pr\left(Q\left(\omega,a\right)=1+1/l\right)= p_{l}\left(a\right) =
B_{a,\gamma}
\frac{\exp(-a)a^{l+\frac{1}{\gamma}}}{\Gamma\left(l+\frac{1}{\gamma}+1\right)}
\]
is a discrete, ''Poisson-like'' distribution with normalization constant 
\[
B_{a,\gamma}=\left(1-\frac{\Gamma(\frac{1}{\gamma},a)}{\gamma \Gamma(1+1/ \gamma)} \right)^{-1}.
\] 
Here $\Gamma(\frac{1}{\gamma},a)$ denotes the incomplete Gamma function:
\[
\Gamma(\frac{1}{\gamma},a)=\int_{a}^{+\infty}\exp(-t)t^{\frac{1}{\gamma}-1}dt
\]
\end{thm}

\begin{proof} Consider random
vector $y=Ax=\left[a^{1/\gamma_{1}}x_{1},\dots,a^{1/\gamma_{n}}x_{n}\right]^{T}$
obtained by multiplying each component $x_{i}$ of $x$ by factor
$a^{1/\gamma_{i}}$ where $a$ is a positive constant. Then
\begin{eqnarray*}
f_{y,q}\left(y\right) & = & \frac{A_{q}}{a^{1/\gamma}}\left(1-\frac{1}{a}\sum_{k=1}^{n}\lambda_{k} \vert y_{k} \vert ^{\gamma_{k}}\right)^{\frac{1}{q-1}}\forall y\in E_{a}
\end{eqnarray*}
with $E_{a}=\left\{ y\in\mathbb{R}^{n}\vert\sum_{k=1}^{n}\lambda_{k}y_{k}^{\gamma_{k}}\le a\right\} $
so that
\begin{eqnarray*}
\sum_{l=0}^{+\infty}p_{l}(a)f_{Ax,1+\frac{1}{l}}(y) & = & B_{a,\gamma}
\sum_{l=0}^{+\infty}
\frac{\exp(-a)a^{l+\frac{1}{\gamma}}}{\Gamma\left(l+\frac{1}{\gamma}+1\right)}
\frac{A_{1+\frac{1}{l}}}{a^{\frac{1}{\gamma}}}
\left(1-\frac{1}{a}\sum_{k=1}^{n}\lambda_{k} \vert y_{k} \vert ^{\gamma_{k}}\right)^{l}
 \\
& =& B_{a,\gamma}  
\left( \prod_{k=1}^{n}\frac{\gamma_{k}\lambda_{k}^{1/\gamma_{k}}}{2\Gamma\left(1/\gamma_{k}\right)} \right)
\sum_{l=0}^{+\infty}
\frac{a^{l}e^{-a}}{\Gamma\left(l+1\right)}
\left(1-\frac{1}{a}\sum_{k=1}^{n}\lambda_{k} \vert y_{k} \vert ^{\gamma_{k}}\right)^{l}
 \\
& =& B_{a,\gamma} 
\left( \prod_{k=1}^{n}\frac{\gamma_{k}\lambda_{k}^{1/\gamma_{k}}}{2\Gamma\left(1/\gamma_{k}\right)} \right)
\exp \left(-\sum_{k=1}^{n} \lambda_{k} \vert y_{k} \vert ^ {\gamma_{k}} \right)
  \\
\end{eqnarray*}
\end{proof}

Figure 3 shows the curves that delimit the north-eastern part of the plane ($\gamma$, $a$) where constant $B_{a,\gamma}$ belongs to the interval $[1-\epsilon, 1]$ with, from bottom to top, $\epsilon=10^{-4}$, $\epsilon=10^{-6}$ and $\epsilon=10^{-8}$. When $(a,\gamma)$ belongs to this area, then the distribution of the random non-extensivity parameter $Q(\omega,a)$ can be approximated accurately as
\be
\label{approxpa}
p_{l}(a) \simeq \frac{\exp(a)a^{l+\frac{1}{\gamma}}}{\Gamma(l+\frac{1}{\gamma}+1)}
\ee
which is again a ``shifted-Poisson'' distribution. Its expectation and variance can then be approximated as follows.

\begin{figure}[htbp]
\begin{center}
\centering
\includegraphics[width= 2.9776in, height= 2.6097in]{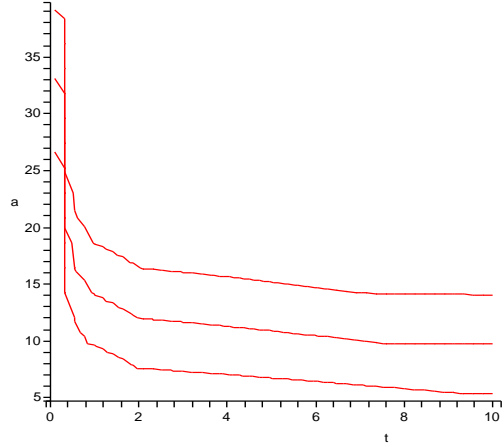}
\caption{areas of the $\gamma,a$ plane where constant $B_{a}$ belongs to $[1-\epsilon, 1]$ with, from bottom to top, $\epsilon=10^{-4}$, $\epsilon=10^{-6}$ and $\epsilon=10^{-8}$ }
\end{center}
\end{figure}

\begin{thm}
The approximated distribution (\ref{approxpa}) has expectation and variance
\be
m=a-\frac{1}{\gamma}, \sigma^{2}=a
\ee
\end{thm}

\begin{proof}
Using approximation (\ref{approxpa}), we get
\be
\nonumber
\sum_{l=0}^{+\infty} l p_{a}(l) = (a-\frac{1}{\gamma}) \left( 1-(\frac{1}{\gamma}-1)\frac{\Gamma(\frac{1}{\gamma}-1,a)}{\Gamma(\frac{1}{\gamma})} \right) +\frac{\frac{1}{\gamma}a^{\frac{1}{\gamma}-1}\exp(-a)}{\Gamma(\frac{1}{\gamma})}
\simeq a-\frac{1}{\gamma}
\ee
and for the variance
\begin{eqnarray*}
\nonumber
\sum_{l=0}^{+\infty} l^2 p_{a}(l) -m^{2}  & = & a+(\frac{1}{\gamma}-a)^{2} +\frac{\Gamma(\frac{1}{\gamma}-1,a)}{\Gamma(\frac{1}{\gamma})} \left( -3\frac{a}{\gamma}+\frac{1}{\gamma^{2}} -\frac{1}{\gamma^{3}} +a+a^{2} +2\frac{a}{\gamma^{2}}-\frac{a^{2}}{\gamma} \right)\\
& - & \exp(-a)a^{\frac{1}{\gamma}-1}(\frac{1}{\gamma^{2}}-\frac{a}{\gamma}+a) -m^{2} \\
& \simeq & a +(\frac{1}{\gamma}-a)^{2} -(\frac{1}{\gamma}-a)^{2}
\end{eqnarray*}

\end{proof}

\subsection{Corollary: Application to an Hamiltonian system}

The result of Theorem 4 can be used in the following framework: let us consider an Hamiltonian system with a phase space of dimension $n=2N$. Denote as $\bold{x}=(q_{1},\dots,q_{N},p_{1},\dots,p_{N})$ the phase space coordinates and consider the following Hamiltonian 
\[
H=\sum_{k=1}^{n} \alpha_{k} \vert x \vert_{k}^{\gamma_{k}}
\]
where $\alpha_{k}$ are positive constants related to the system. Let us assume that the system has maximum Tsallis entropy with parameter $q>1$ under the single energy 
constraint $\langle H \rangle = E$. Its probability distribution $f_{X}$ writes thus
\begin{equation}
\label{fHamilton}
f_{X}(\bold{x})= \frac{1}{Z_{q}} \left( 1 - (q-1)\beta \sum_{k=1}^{n} \alpha_{k} \vert x \vert_{k}^{\gamma_{k}}\right)_{+}^{\frac{1}{q-1}}
\end{equation}
We remark that this distribution coincides with distribution (\ref{eq:tsallis}), except that constants $\lambda_{k}$ in (\ref{eq:tsallis}) are the $n$ Lagrange parameters, whereas there is only one Lagrange parameter in (\ref{fHamilton}), namely the inverse temperature $\beta$. However, results of Theorem 4 apply in the straightforward following way.

\begin {thm}
The Hamiltonian system with probability distribution (\ref{fHamilton}) can be approximated (in the sense of Theorem 4) as a Boltzmann system with distribution
\[
g_{X}(x)=Z_{\beta,q}^{-1}\exp\left(-(q-1)\beta\sum_{k=1}^{n}\alpha_{k} \vert x_{k}\vert^{\gamma_{k}}\right)
\]
with 
\[
Z_{\beta,q}^{-1}=
\left(\prod_{k=1}^{n}\frac{\gamma_{k}\alpha{k}^{1/\gamma_{k}}}{2\Gamma\left(1/\gamma_{k}\right)}\right)
\beta^{\frac{1}{\gamma}}(q-1)^{\frac{1}{\gamma}}
\]
and where the non-extensivity parameter $q$ is a discrete random variable following the shifted-Poisson distribution (\ref{approxpa}).
\end {thm}

\section{Discussion}

\nd We have seen that a Tsallis probability distribution (PD) $S_{q,\sigma^{2}}$ of fixed variance
\be \label{5} S_{q,\sigma^{2}}\left(x\right)=A_{q}\left(1-\frac{x^{2}}{\sigma^{2}}\right)^{\frac{1}{q-1}},\ee
where the normalization constant is given by (\ref{2}), in conjunction with a shifted Poisson distribution
$p(a)$ can be used to expand a Gaussian PD (or more generally, exponential PDs) of fixed variance in the fashion (for brevity we concentrate here upon the Gaussian PD only)

\be \label{6} \frac{1}{\sigma\sqrt{2\pi}}\exp{\left(-x^{2}/2\sigma^{2}\right)} =
\sum_{k=0}^{+\infty}p_{k}\left(a\right)\frac{1}{\sigma\sqrt{2a}}S_{1+1/k,1}\left(\frac{x}{\sigma\sqrt{2a}}\right),  \ee
\nd $\forall x\in\left[-\sigma\sqrt{2a},+\sigma\sqrt{2a}\right].$ Now, it is known that for a Gaussian PD  the variance $\sigma^2$ is proportional to the temperature $T$ (for a purely exponential PD, the temperature squared) \cite{reif}. Thus, Eq. (\ref{6}) entails that, for fixed temperature $T$,   a Gaussian PD (a purely exponential PD) can be thought of as a superposition of Tsallis' PDs of the same temperature. The weights in such an expansion are given by a shifted Poisson PD. Comparing
Eq. (\ref{6}) with Eq. (\ref{bc}) we detect a striking similarity. Of course, in the first we have a ``finite" convolution. Moreover, the roles of Gaussian (purely exponential) and Tsallis PD's are inverted.

\subsection{Finite thermal baths}

The conventional Gibbs' road to derive the canonical distribution \cite{pathria} considers a system $\mathcal{S}$ with energy levels denoted by $\epsilon_i$, weakly interacting with an  infinite thermal bath $\mathcal{B}$ and assume that one describes the ``total" system $\mathcal{T}=\mathcal{S} + \mathcal{B}$ by recourse to Gibbs' {\it micro}canonical ensemble when its total energy $E$ lies in the interval \be \label{7} E_0-\Delta <E< E_0+\Delta,\ee
with $\Delta \ll E_0$ \cite{pathria}.   Plastino and Plastino (PP) \cite{fromgibbs} traverse a new road by assuming that $\mathcal{B}$ is a {\it finite} system, of finite energy $E_b$ and finite number of particles $N$, and in so doing are able to derive, repeating Gibbs' original arguments, {\it not} Gibbs' canonical distribution but that of Tsallis' (Cf. Eqs. (\ref{qestre})-(\ref{qestre1}))

\be \label{8} p_i=\frac{[1-\beta(1-q^*)\epsilon_i]^{1/(1-q^*)}}{Z_{q^*}};\,\,\,Z_{q^*}=\sum_i [1-\beta(1-q^{*})\epsilon_i]^{1/(1-q^{*})}.\ee Assuming further that, plausibly enough, the bath' number of states per unit energy interval grows as a power $\alpha=N$ of $E$ (a common ocurrency \cite{fromgibbs}), PP
 show that the nonextensivity index $q^*$ is given in terms of $N$ in the fashion

 \be \label{9} \frac{N-2}{N-1}\,\,\Longrightarrow \,\,q^* \rightarrow 1\,\,as\,\,both\,\,N,\,\,and \,\,E_0  \rightarrow \infty\,\,with\,\,(N/E_0)=const. \ee Using Eq. (\ref{qestre1}) we have then

 \be \label{10} q=\frac{N}{N-1},  \ee as prescribed by the theorems of the preceding Section.

 \nd We can now interpret the contents of Eq. (\ref{6}) in terms of decomposing the Gaussian PD in terms of Tsallis' ones. In order for thermal equilibrium to be established (and have a Gaussian PD) one  requires, of course, thermal contact with an infinte bath. This reservoir, in turn, can be thought of  as a superposition of finite baths containing different numbers $N$ of particles. Thermal equilibrium with any of them leads to a Tsallis' PD.

 \subsection{Why quasi-Poisson weights in (\ref{6})?}

 \nd A possible visualization of our prevailing state of affairs state is as follows. Due to the short range of the effective
interactions between our system and the bath (e.g., Lennard-Jones, van der Waals, etc.), even if the reservoir is indeed of infinite size the system can not ``appreciate" such a feature. During a short time interval $\Delta T$, the system actually interacts with a finite number of particles, say $M$ of them. Consequently, we may say that the system only ``sees" an effective, finte heat bath of size $M$. Of course, $M$ is constantly changing, due to the
interaction itself and to the time-evolution. Thus our system is constantly ``choosing" $M$ particles to interact with out of a
uniform distribution of them in configuration space. Think, for instance, that the reservoir is an ideal gas in a very large box,
and that the system, during the interval $\Delta T$, interacts with those particles contained within an effective volume $\Delta
V$. Therefore, $M$ would be the number of particles contained within $\Delta V$. But this is a typical situation leading to a
Poisson distribution (for the random variable $M$). Within the present scenario we may say that the Gibbs statistics (infinite
heat bath) arises in an effective way from the superposition (with appropriate weights) of different $q$-statistics (finite heat
baths). In a sense, this situation is the reciprocal of the one considered by Beck \cite{BC03,B04a,B01}. In Beck's formulation,
the $q$-statistical ensemble constitutes an effective description of a system interacting with a heat bath of fluctuating
temperature. Here we show that, under physically sensible assumptions, it is possible to regard the Gibbs statistics as
describing a system interacting with a finite heat bath of fluctuating size.

It is interesting to examine this mechanism from the point of view
of the ergodic hypothesis. According to this hypothesis, ensemble
averages of dynamical quantities coincide with suitable  time
averages \cite{pathria}. Now, if our $q$-based origin of the
Gibbs' statistics is actually realized for a given system, we
should expect the ergodic condition to hold ``differently" in two
distinct time scales. (i) In the first place, time averages
computed during appropriate, finite time intervals $\Delta T$
would coincide with $q$-averages, with the $q$-values associated
with successive $\Delta T$-time windows distributed in a
Poisson-like way. (ii) Secondly, time averages computed during
time intervals $\tau$ much longer than $\Delta T$, would coincide
with Gibbs' ensemble averages. This kind of behavior has some
similarities, as well as important differences, with an scenario
leading to the $q$-statistics that has been suggested by Tsallis'
\cite{gellmann}. According to this Tsallis' picture, some systems
with long-range interactions relax first to a meta-stable state
(characterized by a $q$-statistics), which is endowed with a long
mean-life. But, eventually, the system relaxes to a final state
described by the Boltzmann-Gibbs' statistics. The mechanism {\it
here}  suggested by us shares with that of Tsallis' the fact of (i)
being based on two different relaxation time scales, and (ii)
recovering the Boltzmann-Gibbs statistics in the limit of long
relaxation times. However, our mechanism differs in one
fundamental way from the Tsallis' one: in our scenario the system
is always described by a $q$-statistics, if observed only during a
relatively short interval $\Delta T$.

Further developments and applications of our present
$q$-fluctuating approach to the Gibbs' ensemble, and its possible
verification in (actual or numerical) experiments, would certainly
be greatly welcome.

\end{document}